\documentclass[prd,twocolumn,showpacs,preprintnumbers,amsmath,amssymb]{revtex4}

\usepackage{graphicx}

\newcommand{\pipi}{$\pi\pi$ }
\newcommand{\fsig}{$f_0(500)$ }
\newcommand{\sig}{$\sigma$ }
\newcommand{\mpipi}{$m_{\pi\pi}$ }

\begin{document}

\title{Modified \mbox{\boldmath $\pi\pi$} amplitude with \mbox{\boldmath $\sigma$} pole}

\author{P. Byd\v{z}ovsk\'y$^a$, R. Kami\'nski$^b$, V. Nazari$^b$}
\affiliation{$^a$Nuclear Physics Institute, ASCR, \v{R}e\v{z}/Prague 25068, Czech Republic\\
$^b$Institute of Nuclear Physics, Polish Academy of Sciences, Krak\'ow 31-342, Poland}


\begin{abstract}

A set of well known once subtracted dispersion relations with imposed crossing 
symmetry condition is used to modify unitary multichannel $S$ ($\pi\pi$, $K \bar K$,  
and $\eta\eta$) and $P$ ($\pi\pi$,  $\rho 2\pi$, and $\rho\sigma$) wave amplitudes 
mostly below 1 GeV. Before the modifications, 
these amplitudes significantly did not satisfy the crossing symmetry condition and 
did not describe the \pipi threshold region. Moreover, the pole of the $S$ wave amplitude 
related with the \fsig meson (former $f_0(600)$ or $\sigma$) had much smaller imaginary 
part and bigger real one in comparison with those in the newest Particle Data Group Tables.
Here, these amplitudes are supplemented by near threshold expansion polynomials and 
refitted to the experimental data in the effective two pion mass from the threshold 
to 1.8~GeV and to the dispersion relations up to 1.1~GeV.
In result the self consistent, i.e. unitary and fulfilling the crossing symmetry 
condition,  $S$ and $P$ wave amplitudes are formed and the \sig pole becomes much 
narrower and lighter. 
To eliminate doubts about the uniqueness of the so obtained sigma pole position 
short and purely mathematical proof of the uniqueness of the 
results is also presented. 
This analysis is addressed to a wide group of physicists and aims at 
providing a very effective and easy method of modification of, many 
presently used, $\pi\pi$ amplitudes with a heavy and broad $\sigma$ meson 
without changing of their original mathematical structure.
\end{abstract}


\maketitle
%
%
\section{Introduction}
\label{Introduction}

New once-subtracted dispersion relations with the imposed crossing symmetry condition 
for the $S$-$F$ wave \pipi amplitudes have recently been derived and 
presented in the Refs. \cite{GKPY,osdrdf}. Further analysis of these equations in 
the Ref. \cite{PreciseDet} for the $S$ and $P$ waves (the so-called GKPY equations) led 
inter alia to very precise determination of the position of the pole related with 
the \fsig resonance (hereafter $\sigma$).

Importance and effectiveness of similar dispersion relations but with two subtractions, 
i.e. of the so-called Roy equations \cite{Roy1971}, were already presented on the example 
of the elimination of long standing up-down ambiguity in the \pipi $S$ wave amplitude 
below 1~GeV \cite{Kaminski:2002,Pennington1975} and of  determination of the quark condensate 
constants \cite{Descontes}.

Quite recently the Roy's equations were once again effectively exploited in searching 
for unique determination of the $S$ and $P$ wave \pipi scattering 
amplitudes~\cite{Bern,A4,Caprini:2005zr}. Due to incorporation of two boundary conditions 
for these amplitudes, it was possible to find such analytical solution below 800~MeV 
in accordance with derived and proven theorem on the uniqueness of such solutions~\cite{Wanders2000}. 
One of the byproduct of these analyses and those with GKPY equations \cite{GKPY,PreciseDet} 
was official and long-awaited significant modification of the position of the \sig pole 
in Particle Data Tables. For many years this state was appearing with mass and width 
noticeably larger than 500~MeV. For example in the Particle Data Tables in 2010~\cite{PDGTables2010} 
the mass was in the range $M =$ 400--1200~MeV and the full width $\Gamma =$ 600--1000~MeV. 
Before year 1994 the \sig meson was even excluded from the Tables for about 20 years.
Since 2012 its parameters are much better determined, i.e., $M =$ 400--550~MeV and 
$\Gamma =$ 400--700~MeV~\cite{PDGTables2012}. The reason for this many years of confusion 
and uncertainty about these parameters was that their determination was based mostly 
on fairly disparate and uncertain experimental results. Fortunately, well-grounded theoretical 
works based on dispersion relations with the imposed crossing symmetry condition, 
presented, e.g. in \cite{GKPY,PreciseDet,Bern,A4,Caprini:2005zr}, provided very strong 
arguments to resolve the existing uncertainties.

Despite of those big and widely accepted changes in parameters of the \sig meson many analyses 
can still use the old, i.e. significantly too wide and too massive, scalar-isoscalar state 
below 1~GeV. The reason for this may be difficulties in changing parameters of some models 
or parametrizations to adapt them to the new requirements.
Use of the correct and precise parametrizations of the \pipi amplitudes is, 
however, sometimes crucial especially when high precision of the final results is required.
This can be particularly well seen, for example, in analyses of the $\pi\pi$ final state 
interactions in the heavy mesons decays (e.g. $B$ or $D \to M\pi\pi$ where $M$ is 
$K$ or $\pi$) needed to determine parameters of a very small CP violation.

Another kind of analyses which need correct and very precise \pipi amplitudes are those 
which pretend to describe spectrum of light mesons decaying into \pipi pairs in given 
partial waves and which strongly require verification of compliance with the crossing 
symmetry condition. One of them is the multichannel ($\pi\pi$, $K \bar K$, and $\eta\eta$) 
analysis of the \pipi scattering data presented in \cite{Yura2010,Yura2012a,Yura2012b} 
which uses unitary amplitudes up to 1.8~GeV with proper analytical properties on the 
whole Riemann surface. However, in the construction of these amplitudes the crossing 
symmetry condition was not required what resulted in insufficiently precise description 
of the \pipi elastic region. Moreover, these amplitudes did not describe correctly 
the experimental data in the vicinity of the $\pi\pi$ threshold. 

The aim of this work is to present a general method of refining the \pipi amplitudes 
by fitting them to the GKPY equations while maintaining their original mathematical 
structure. The method alters only numerical values of some of parameters (e.g. positions 
of the resonances and the background). The refined amplitudes must be self-consistent, 
i.e. still unitary and, what is new, fulfilling the crossing symmetry condition below 1.1~GeV.
The method is demonstrated on a refining of the multichannel $S$  and $P$ wave amplitudes 
from \cite{Yura2010}. The insufficiency of these amplitudes to describe the near threshold 
phase shifts complicates slightly the refining since a new prescription of the near 
threshold amplitudes has to be added. This makes, however, the presented method more 
general and enhances  a chance to convince potential readers, who might want to perform 
similar refining of their amplitudes, on effectiveness and simplicity of the proposed 
method. Quantitative changes in the refined amplitudes will be presented inter alia 
by the difference between the positions of the $\sigma$ poles before and after fitting.

In Sec. II we recall the method used in the multichannel analysis \cite{Yura2010} 
and its main results. In Sec. III we present the dispersion relations with the imposed 
crossing symmetry condition and methodology used in dispersive analyses of the 
data~\cite{GKPY}. In Sec. IV we provide more details of the analysis and present 
results of refining the amplitudes, constructed in \cite{Yura2010} and supplemented 
with a near threshold part, fitting them to the data and to the dispersion relations.
Discussion of results is completed in Sec. V in which we demonstrate a simple proof 
of correctness and uniqueness of the results. Conclusions are drawn in the last Section.

%
%
\section{Multichannel \mbox{\boldmath $S$} and \mbox{\boldmath $P$} wave amplitudes}
\label{MultichannelSP}
The three-channel unitary S-matrix is constructed using the method 
which has been previously applied to analyses of the multichannel 
$\pi\pi$ data~\cite{Yura2010,Yura2012a,Yura2014}. 
The method is based on the uniformizing variable and the formulas 
for proper analytical continuation of the S-matrix elements to all 
sheets of the Riemann surface. More details on the method and notation 
can be found in Refs.~\cite{Yura2010} and \cite{Yura2012a}. Here we 
give only basic principles and formulas needed in calculations 
of the amplitudes.

Two channels ($K\bar{K}$, $\eta\eta^\prime$)  coupled to the $\pi\pi$ 
channel for the $S$ wave and ($\rho 2\pi$, $\rho\sigma$) for the $P$ wave 
are explicitly included. The eight-sheeted Riemann 
surface is transformed into a simpler complex plane utilizing 
the uniformizing variable 
\begin{equation}
\label{uniform_var}
w= \frac{\sqrt{s-s_2} + \sqrt{s-s_3}}{\sqrt{s_3-s_2}},
\end{equation}
where $s$ is the squared effective two-pion mass ($s=m_{\pi\pi}^2$) and 
$s_2$ and $s_3$ are thresholds of the second and third channel, 
respectively. The transformation into the uniformizing plane can 
be fully done if and only if less than three-channels are considered. 
In the three-channel 
case we have to neglect influence of the lowest $\pi\pi$ branch point 
keeping, however, unitarity on the $\pi\pi$ cut to construct a 
four-sheeted model of the initial Riemann surface. This approximation 
allows us to get a simple description of multichannel resonances 
via seven types of pole clusters on the uniformizing plane denoted as 
a, b, c,..., g, (for more details see Refs.~\cite{Yura2010} and \cite{Yura2012a}). 
Neglecting the  $\pi\pi$ branch point, however, leads
to a bad description of the data near the $\pi\pi$ 
threshold~\cite{Yura2010,Yura2012a,Yura2014}. 
Note that, in the two-channel case this approximation is not needed 
and the threshold data are described correctly~\cite{Yura2012b,Yura2008}.
 
In Eq.(\ref{uniform_var}) the left-hand branch point connected 
with the $t$ channel is not taken into account which  
means that the crossed channels are not explicitly considered 
in the construction of the amplitudes. A contribution of the left-hand 
cut is, however, included in the background part of the amplitude. 
Note that, in Refs.~\cite{Yura2012a} and \cite{Yura2014} the left-hand 
branch point in $w$ was already included in the $S$ wave analysis. 
In the presented analysis we use the three-channel formalism 
of Ref.~\cite{Yura2010} because it was consistently applied 
both to $S$  and $P$ wave amplitudes and the crossing symmetry restoration 
effect is expected to be larger in this case.   

The S-matrix elements of all assumed coupled processes are 
expressed in terms of the Jost matrix determinant ($d$) using 
the Le Couteur-Newton relations~\cite{Yura2010}. They are taken 
as products of the resonant ($S^{res}$) and background ($S^{bgr}$) 
parts anticipating that the main effect of the resonances 
is given by the pole clusters in $S^{res}$ and small remaining 
contributions of resonances and neglected singularities 
(e.g. the left-hand branch point) can be included via $S^{bgr}$.
The $\pi\pi$ S-matrix element reads as
\begin{equation}
\label{s11_element}
S_{11}= S^{res}_{11} S^{bgr}_{11} =  \frac{d^*_{res}(-w^*)}{d_{res}(w)} 
\frac{d_{bgr}(-k_1,k_2,k_3)}{d_{bgr}(k_1,k_2,k_3)},
\end{equation}
where $k_j$ are the channel momenta. The $d$ function for 
the resonant part has a simple polynomial-like form 
\begin{equation}
\label{d_res}
d_{res}(w) = w^{-M/2}\prod_{r=1}^M (w+w_r^*),
\end{equation}
where $M$ is a number of all poles of the $S_{11}$ at $w = -w_r^*$ 
related with resonances. Various scenarios, which differ in the number 
and the types of resonances, were fitted in \cite{Yura2010} to the 
experimental data and the most probable one (with the smallest 
$\chi^2$) was selected. Note that, after the number and poles of 
the resonances are fixed by fitting to the data, the resonance part 
of the S matrix is known in the whole Riemann surface free of 
uncertainties due to its analytical continuation to the non physical 
region~\cite{PreciseDet,Caprini,Caprini:2005zr}. 
 
The background part is modeled in the physical region via complex 
energy-dependent phases to mimic the opening of the channels whose 
branch points are not included in the uniformizing variable.
In the $S$ wave the $d$ function for the background is
\begin{equation}
\label{bgr_s-wave}
d_{bgr}(k_1,k_2,k_3) = \exp\left[-i\sum_{n=1}^3 
\frac{k_n}{m_n}(\alpha_n+i\beta_n) \right], 
\end{equation}
where 
\begin{eqnarray} 
\alpha_n &=& a_{n1} + 
a_{n\sigma}\frac{s-s_\sigma}{s_\sigma}\theta(s-s_\sigma)+ 
a_{nv}\frac{s-s_v}{s_v}\theta(s-s_v)\nonumber\\
& & + a_{n\eta}\frac{s-s_\eta}{s_\eta}\theta(s-s_\eta),\\ 
\beta_n &=& b_{n1} + 
b_{n\sigma}\frac{s-s_\sigma}{s_\sigma}\theta(s-s_\sigma)+ 
b_{nv}\frac{s-s_v}{s_v}\theta(s-s_v) \nonumber\\
& & + b_{n\eta}\frac{s-s_\eta}{s_\eta}\theta(s-s_\eta), 
\end{eqnarray} 
and $s_\sigma$ and $s_\eta$ are the $\sigma\sigma$ and $\eta\eta$ 
thresholds, respectively. 
An effective threshold describing influence of 
many opened channels in the vicinity of 1.5~GeV 
(e.g., $\rho\rho,~\omega\omega$) is denoted by $s_v$ and together with 
the $s_\sigma$ is determined in the analysis. 
The $m_n$ denotes mean channel masses.
In the $P$ wave the background is 
\begin{equation} 
\label{bgr_p-wave} 
 d_{bgr}(k_1,k_2,k_3) = \exp\left[-ia\;{\rm sign}(k_1) + 
 b\left(\frac{k_1}{m_1}\right)^3\right], 
\end{equation} 
where $a$ and $b$ are real numbers. 

The resonance poles $-w_r^*$ and background parameters in Eqs.(\ref{bgr_s-wave}) 
and (\ref{bgr_p-wave}) were obtained, in Ref. \cite{Yura2010} from fitting 
the phase shifts and inelasticity parameters in the assumed channels to 
experimental data. 
The scattering $\pi\pi$ amplitude in a given partial wave ($\ell$) and 
isospin ($I$) is related to the corresponding S-matrix element
\begin{equation} 
\label{partial_amplitude} 
 t_{\ell}^{I}(s) = \frac{\sqrt{s}}{2\,k_1}\;\frac{(S_{11})^I_\ell-1}{2i}\,.
\end{equation} 

\begin{figure}[h!]
\includegraphics[scale=0.33]{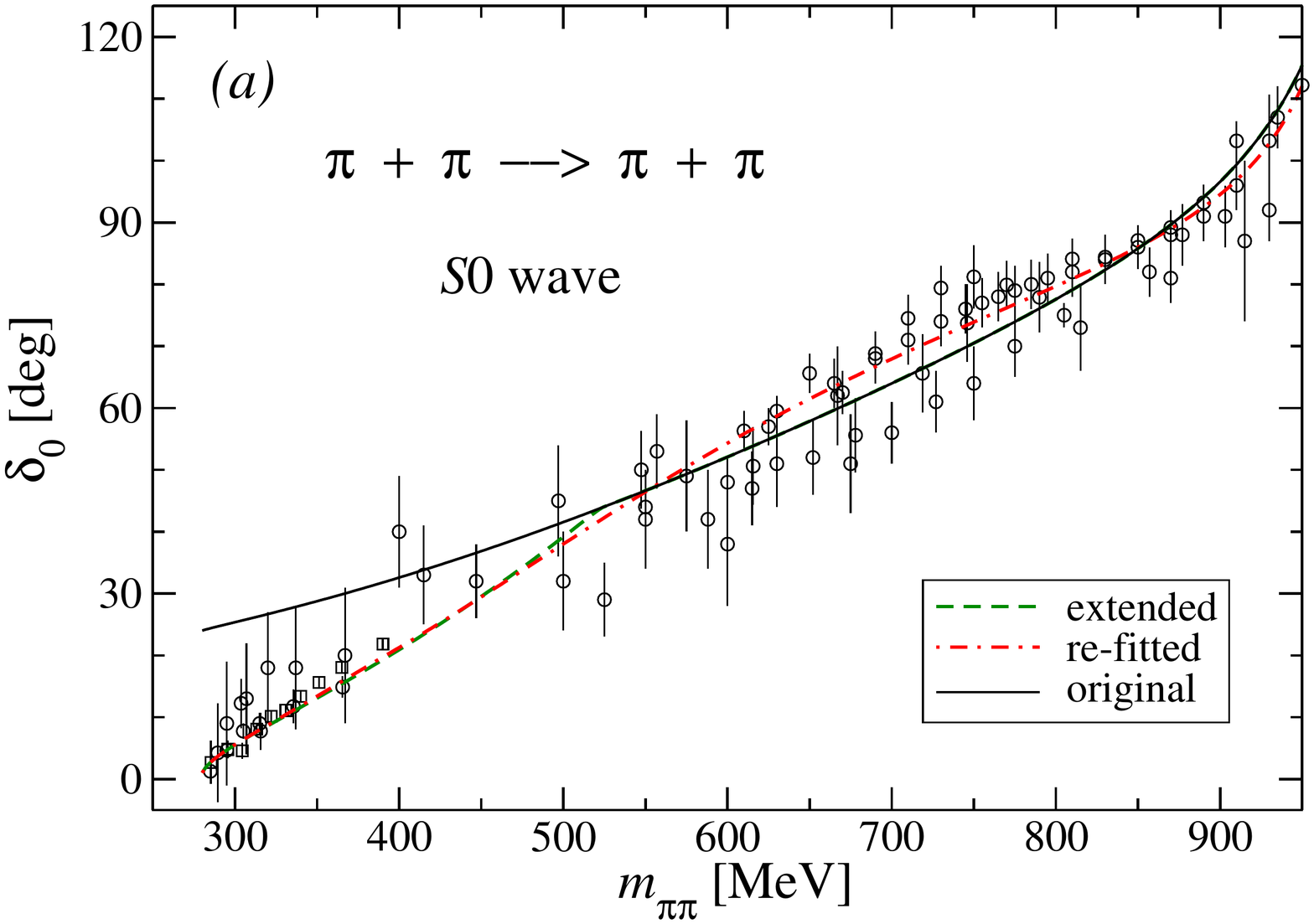}\\
\includegraphics[scale=0.33]{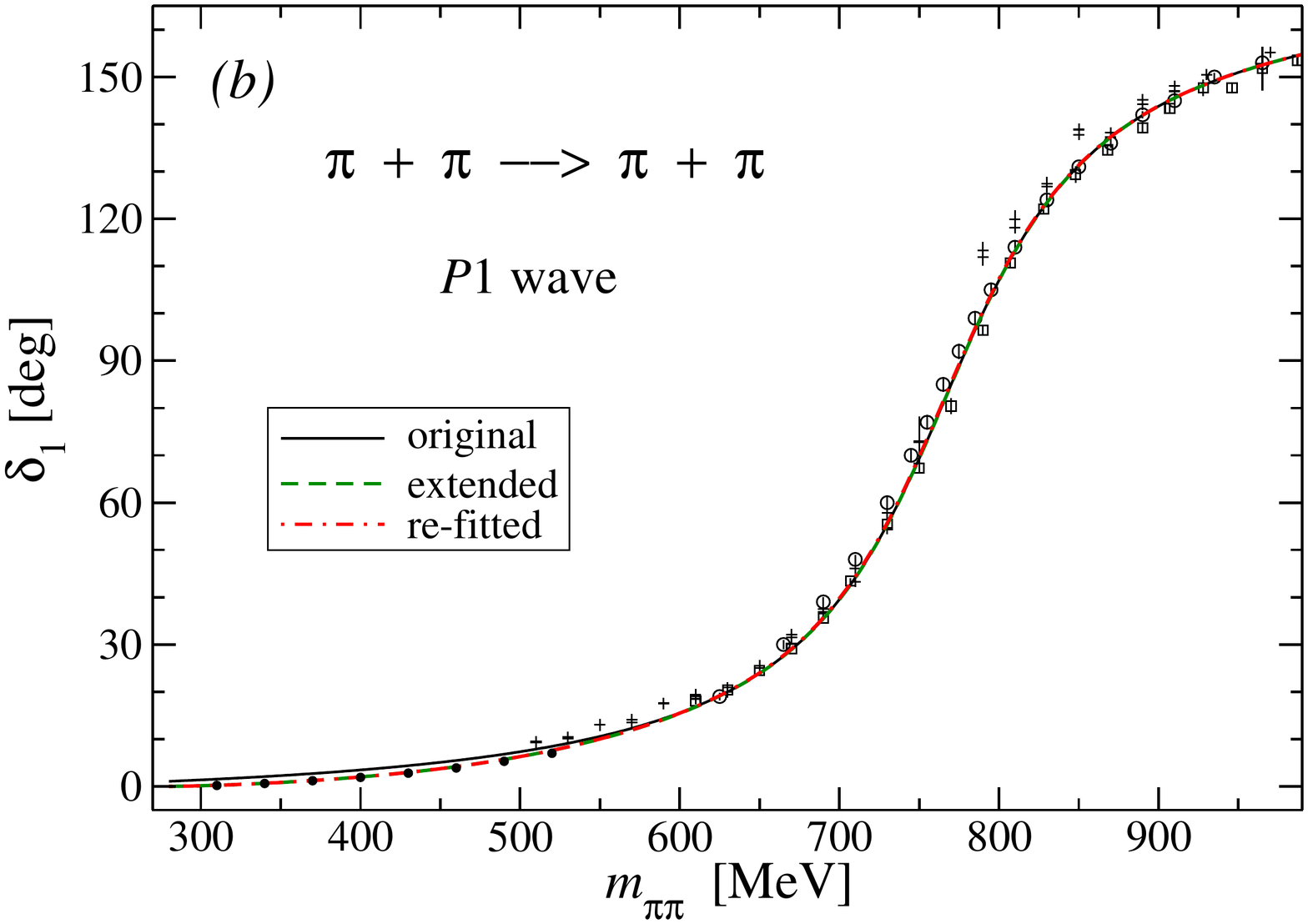}
\caption{Phase shifts for the $S0$ wave ($a$) and $P1$ wave ($b$) 
amplitudes as a function of the effective two pion mass $m_{\pi\pi}$ 
for the original (solid line), extended (dashed line) and refitted 
(dash-dotted line) amplitudes considered in the text. 
Data are taken from \cite{Yura2010}.}
\label{FigBydzNoDR}
\end{figure}

\begin{figure}[h!]
\includegraphics[scale=0.34]{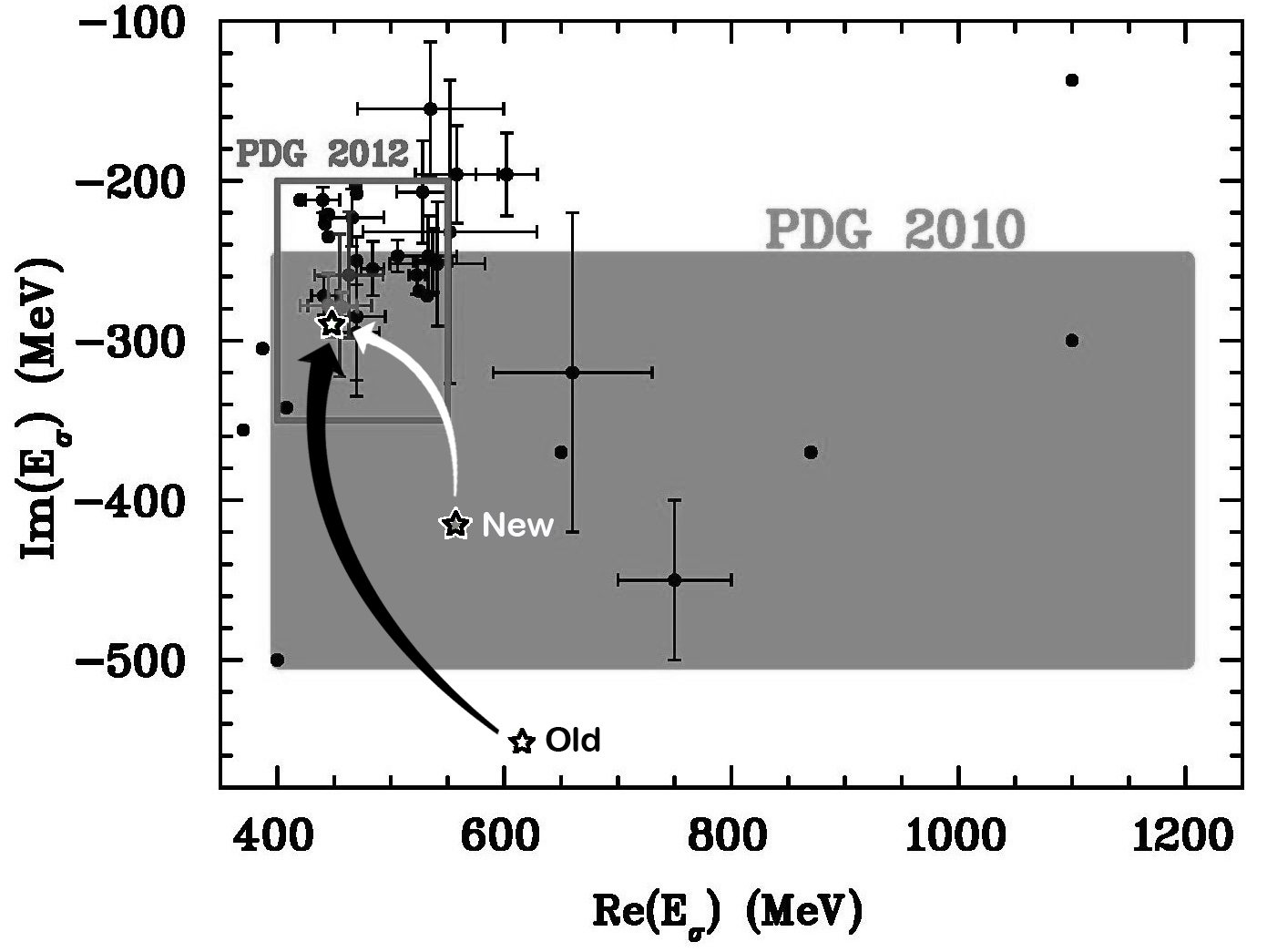}
\caption{Shift of the $\sigma$ pole on sheet II after fitting to the GKPY 
equations. The stars in the lower part of figure denote the pole positions 
for the ``Old" (original) and ``New" amplitudes while the stars indicated by 
the arrows their positions after refitting. The big (PDG2010) and small 
(PDG2012) rectangles show area allowed by the Particle Data Tables in 2010 
and 2012, respectively. Black points are positions of the poles listed in 
the Particle Data Tables published in 2010.}
\label{fig:InOut.D0}
\end{figure}

Mainly because in the presented analysis we aim to demonstrate our method,  
we used the amplitudes from Ref.~\cite{Yura2010}: variant II for the $\ell I = S0$ 
wave ($\ell =0, I=0$), i.e., the scenario in which the resonances 
$f_0(500)$, $f_0(1370)$, $f_0(1500)$ and $f_0(1710)$ are described by 
clusters of type a, b, d and c, respectively and the $K\bar{K}$-threshold 
resonance 
$f_0(980)$ is represented only by the pole on sheet II and shifted pole 
on sheet III. In the $P1$ wave ($\ell =1, I=1$) we have chosen the best 
scenario (see ``aebbc" 
in Table III in \cite{Yura2010}) in which the vector resonances $\rho(770)$, 
$\rho(1250)$, $\rho(1450)$, $\rho(1600)$ and $\rho(1800)$ are described 
by the clusters of type a, e, b, b and c, respectively.
These amplitudes describe satisfactorily the data in all assumed channels 
up to 1.8~GeV except for the $\pi\pi$ phase shift in the $S0$ wave below 
500~MeV, see Fig.~\ref{FigBydzNoDR} for the ``original" amplitude, which 
is due to neglecting the $\pi\pi$ branching point. 
The position of the $\sigma$ pole on sheet II of the Riemann surface 
is quite far from the values recommended by the Particle Data Tables 
(see the star denoted by ``Old" in Fig.~\ref{fig:InOut.D0}) which is 
attributed to the absence of the crossing-symmetry constraints in 
constructing the multichannel amplitudes. 
However, the pole of the $\rho$(770) is in a proper position 
giving the correct mass and width~\cite{Yura2010}. These results show 
the importance of the crossing symmetry in the $S0$ wave below 800~MeV. 
Note that, in the multichannel analysis the partial waves are treated 
fully independently and the crossing symmetry, being applied to the full 
amplitude, introduces correlations between the parameters in the 
$S0$  and $P1$ waves.

%
%

%
%

%
%
\section{Crossing symmetry constraints imposed on the amplitudes}
\label{S0P1amplitudes}

In Ref. \cite{GKPY} it has been presented that, due to only one subtraction, 
the GKPY equations are much more demanding, i.e. have significantly smaller 
uncertainties than the Roy equations with two subtractions. Therefore in this 
analysis we use only the GKPY equations which have the following general form
\begin{eqnarray}\nonumber 
{\mbox{Re } \bar t_{\ell}^{I}(s)} & = & \sum\limits_{I'=0 }^2 C^{II'}{ t^{I^\prime}_0(4m_{\pi}^2)} \\  
& + &  
\displaystyle \sum\limits_{I'=0}^2
        \displaystyle \sum\limits_{\ell'=0}^3
     \hspace{0.2cm}-\hspace{-0.48cm}
        \displaystyle \int \limits_{4m_{\pi}^2}^{\infty}\ \hspace{-0.2cm}ds'
     K_{\ell \ell^\prime}^{I I^\prime}(s,s')\, { \mbox{Im }t_{\ell'}^{I^{\prime}}(s')},
\label{EqGKPY}
\end{eqnarray}
where $t_{\ell^{\prime}}^{I^{\prime}}(s')$ and $\bar t_{\ell}^{I}(s)$ are the 
input and output amplitudes, respectively, in a given partial wave $\ell, \ell^\prime$ 
with isospin $I, I^\prime$. The $C^{II'}$ is the crossing  matrix constant and 
$K_{\ell \ell^\prime}^{I I^\prime}(s,s')$ are kernels constructed for partial 
wave projected amplitudes with the imposed $s \leftrightarrow t$ crossing 
symmetry condition. Given $t^I_{\ell}(s)$ amplitude fulfills the crossing 
symmetry when real part of the output amplitude ${\mbox{Re } \bar t_{\ell}^{I}(s)}$ 
is equal to real part of the input one  ${\mbox{Re } t_{\ell}^{I}(s)}$. 
In practice the part of Eq. (\ref{EqGKPY}) containing sums of the integrals 
is divided into two components. The first one contains contributions from lower 
energy parts (i.e. for $\sqrt s < 1.42$~GeV) and is called ``kernel part'' while 
the second one includes contributions from amplitudes at higher energies and 
is called ``driving terms''. Full expressions for the GKPY equations, together 
with derivation can be found in~\cite{GKPY}.
The derivation starts with the Cauchy theorem applied to the full \pipi amplitude 
depending on the $s$ and $t$ Mandelstam variables and composed of a set of partial 
waves afterwards integrated over $t$.
Of course the Cauchy theorem can not be arbitrary used to a single partial
wave without any physical constrain, which in this case is just the crossing symmetry 
relating all partial waves together.

In order to check how the $S0$ and $P1$ wave amplitudes described in 
Sec.~\ref{MultichannelSP} fulfill the crossing symmetry condition one has to use them 
as the input amplitudes in Eq.~(\ref{EqGKPY}), i.e. integrate their imaginary parts 
with proper kernels from the threshold to 1.42~GeV. However, as we have already 
explained, both these amplitudes suffer from an improper description of the phase 
shifts near the $\pi\pi$ threshold region. Therefore, in order to allow the integration 
from the \pipi threshold we have re-defined these amplitudes for energies 
$\sqrt{s_1} < \sqrt{s} < \sqrt{s_{0\ell}}$ where $\sqrt{s_{00}}$ and $\sqrt{s_{01}}$ 
are matching energies for the $S0$ and $P1$ waves, respectively. Above these energies 
the amplitudes remain fully equivalent to the ``original" multichannel ones 
from Sec.~\ref{MultichannelSP}. Below these energies the amplitudes are 
parametrized as     
\begin{eqnarray}\nonumber 
\label{EqThreshold}
\sin\,2\delta^I_\ell & = & \frac{4m_{\pi}k_1^{2\ell+1}}{\sqrt s} \\ 
& \times & \left[a^I_\ell + b^I_\ell k_1^2 + c^I_\ell k_1^4 + d^I_\ell k_1^6 + 
\mathcal{O}(k_1^8)\right], 
\end{eqnarray}
where $\delta^I_\ell$ denotes the phase shift in the $\ell I$ wave.  
The parameters $a^I_\ell$ and $b^I_\ell$ are the scattering length and the so-called 
slope parameter, respectively, which can be fixed or fitted to the data 
and to the dispersion relations. In this analysis they were fixed at the values: 
$a^0_0=0.220$~$m_\pi^{-1}$, $a^1_1=0.0381$~$m_\pi^{-3}$, $b^0_0=0.278$~$m_\pi^{-3}$, 
and $b^1_1=0.00523$~$m_\pi^{-5}$ following the results of \cite{GKPY}.  
The parameters $c^I_\ell$ and $d^I_\ell$ are used to match smoothly the  
phase shifts (\ref{EqThreshold}) with the multichannel ``original" ones from 
Sect.~\ref{MultichannelSP} at the matching energies $\sqrt{s_{0\ell}}$. 
They are therefore calculated from the conditions that 
the phase shifts and their derivatives are continuous functions at $\sqrt{s_{0\ell}}$.
Note that, one can also use other parametrizations in Eq.~(\ref{EqThreshold}), 
e.g. the generalized effective range expansion or the form used in Ref.~\cite{Bern}.

Hereafter the $\pi\pi$-threshold-region corrected original amplitudes are denoted 
by ``extended" amplitudes. In Fig.~\ref{FigBydzNoDR} the phase shifts of these 
amplitudes are presented as the dashed lines. Values of the matching energies $\sqrt{s_{00}}$ 
and $\sqrt{s_{01}}$ were fitted to the phase shifts to achieve the best description of the 
data. In Fig.~\ref{FigBydzNoDR}, the values were found to be 525 and 643~MeV for 
the $S0$ and $P1$ amplitudes, respectively. Above these matching energies the original 
and extended amplitudes are equivalent and below they follow the expansions 
(\ref{EqThreshold}). One can see that the phase shifts of both extended amplitudes 
follow quite well the data below and above the matching energies.

\begin{figure}[h!]
\centerline{\includegraphics[angle=0,scale=0.45]{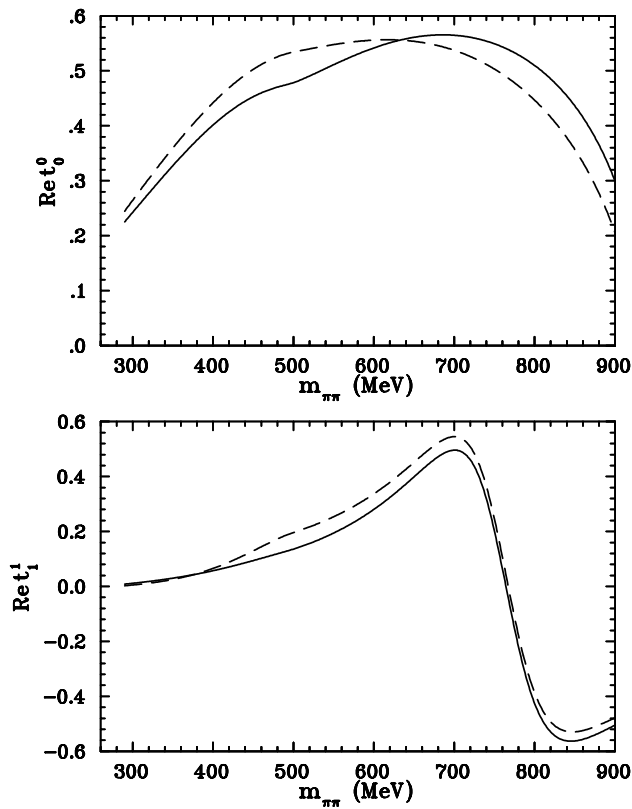}}
\caption{Input (dashed line) and output (solid line) real parts of the $S0$ wave 
(upper figure) and $P1$ wave (lower figure) extended amplitudes, i.e. before fitting.}
\label{FigReOld}
\end{figure}

Figure \ref{FigReOld} illustrates that these extended amplitudes, however, 
do not fulfill the crossing symmetry condition especially for the $S0$ wave.
A big difference is very well seen between the real parts of the input and 
output amplitudes, particularly below about 600~MeV. Also in the case of the 
$P1$ wave  we see worse agreement between the input and the output amplitudes 
in comparison with that in Fig. 13 in \cite{GKPY} where all partial waves 
have been fitted inter alia to the GKPY equations. 
The saddle point in the $S0$ output amplitude near the 
500~MeV is caused by a noticeable change of curvature 
of the phase shifts seen in Fig. \ref{FigBydzNoDR} near the matching energy.
%
%

%
%
\section{Results of refining}
\label{Results}

To improve agreement of the $S0$  and $P1$ wave amplitudes with the crossing 
symmetry, the corresponding extended amplitudes have been fitted to the GKPY 
dispersion relations (hereafter DR) and to the data.
Taking advantage of the fact that these equations also apply to the $S2$ 
amplitude, the output of which strongly depends on the input amplitude $S0$, 
the $S2$ has also been fitted to the DR.
The $S2$ input amplitude together with those for the $D0$, $D2$ and $F1$ 
has been taken from \cite{GKPY} and fixed.  

The total $\chi^2$ was composed of five parts 
\begin{equation}
\label{EqChi2Total}
\chi^2 = \sum_{l=1}^{2}\chi^2_{Data}(l) + \sum_{l=1}^{3}\chi^2_{DR}(l)
\end{equation}
where $l = 1, 2, 3$ itemizes the $S0$, $P1$ and $S2$ partial waves, respectively.
Corresponding $\chi^2_{Data}(l)$ and  $\chi^2_{DR}(\l)$ are expressed by
\begin{equation}
\label{EqChi2Data}
\chi^2_{Data}(l) = 
\sum_{i=1}^{N_{\delta}^l} \frac{(\delta_i^{exp}-\delta_i^{th})^2}
{(\Delta \delta_i^{exp})^2} + 
\sum_{i=1}^{N_{\eta}^l} \frac{(\eta_i^{exp}-\eta_i^{th})^2}
{(\Delta \eta_i^{exp})^2}
\end{equation}
and
\begin{equation}
\label{EqChi2DR}
\chi^2_{DR}(l) = \sum_{i=1}^{N_{DR}} \frac{\left[{\rm Re}\:\bar t_{\ell}^{I}(s_i)-
{\rm Re}\:t_{\ell}^{I}(s_i)\right]^2}{\left[\Delta {\rm Re}\:\bar t_{\ell}^{I}(s_i)\right]^2}.
\end{equation}
The experimental phase shifts $\delta_i^{exp}$ and inelasticities $\eta_i^{exp}$ 
for a given partial wave $\ell I$ (in all considered channels and with 
corresponding errors) are those used in \cite{Yura2010} for \mpipi 
above 0.6~GeV and 0.992~GeV, respectively. 
Apart from these data sets new data from  \cite{ThrData} for the near threshold 
region have been used. They have been obtained in experiments for $K_{l4}$ decays.
Theoretical values of the phase shifts $\delta_i^{th}$ and inelasticities 
$\eta_i^{th}$ have been calculated from the multichannel amplitudes and from 
the threshold expansion (\ref{EqThreshold}) above and below the matching energy, 
respectively. The output amplitudes ${\rm Re}\,\bar t_{\ell}^{I}(s_i)$ 
in (\ref{EqChi2DR}) are calculated using the GKPY equations (\ref{EqGKPY}) 
and their errors are fixed to 0.01 in order to make the 
$\chi^2_{DR}=\sum \chi^2_{DR}(l)$ part of the total $\chi^2$ comparable 
with the $\sum \chi^2_{Data}(l)$. The input amplitudes ${\rm Re}\,t_{0}^{0}(s_i)$ 
and ${\rm Re}\,t_{1}^{1}(s_i)$ come directly from the extended amplitudes. 
The total number of the data points 
$N_{\delta}^{S0} + N_{\eta}^{S0} + N_{\delta}^{P1} + N_{\eta}^{P1}$ is 494 
while $N_{DR}$ for DR was chosen to be 26 for each fitted partial wave to cover 
the \mpipi range from 0.31~GeV to 1.09~GeV with step 0.03~GeV.

In the fitting we changed only those parameters of the $S0$ and $P1$ extended 
amplitudes which can strongly influence the elastic region.
The free parameters considered in the $S0$ wave are: the background parameters 
in the elastic channel $a_{11}$, $a_{1\sigma}$, $a_{1v}$, $a_{1\eta}$, $b_{1\sigma}$, 
$b_{1v}$ and $b_{1\eta}$, the matching energy $\sqrt{s_{00}}$ and poles of the resonances 
$f_0(500)$, $f_0(980)$ and $f_0(1500)$. The resonance $f_0(1500)$ was included because 
it was found to contribute significantly to the elastic phase shift below 1~GeV. 
Note that, in our previous analysis in Ref. \cite{FB20} this resonance was not included.
In the $P$ wave only the background parameters $a$ and $b$, the matching energy
$\sqrt{s_{01}}$ and the $\rho(770)$ were included. Total number of free parameters, 
i.e. those in the resonant and background parts in Eq.~(\ref{s11_element}), is 31.

In Ref.~\cite{Yura2010} the number and values of the fitted resonance parameters
were restricted assuming a simple Breit--Wigner parametrization and some 
constraints imposed on positions of the poles. It ensured a compactness 
of resonance clusters and simultaneously reduces a number of free parameters: 
four for the $f_0(500)$, $f_0(980)$ and $\rho(770)$ and eight for the $f_0(1500)$. 
Following this assumption we refitted the considered free parameters of the $S0$ 
and $P1$ extended amplitudes. In Table \ref{TableChi2-1} we show values of the full 
$\chi^2$ and its contributions from the $S0$ and $P1$ amplitudes and the DR before 
(extended) and after (refitted) fitting. 
%
%
\begin{table}[h!]
\hspace{.0cm}
\begin{tabular}{lcccc}  
\hline
& \multicolumn{1}{c}{$\chi^2$} & \multicolumn{1}{c}{$\chi^2_{Data}(S0)$} & 
\multicolumn{1}{c}{$\chi^2_{Data}(P1)$} & \multicolumn{1}{c}{$\chi^2_{DR}$} \\
\hline
extended  & 1122.5  & 339.4 & 305.1 & 478.0 \\
refitted & $\>\>$639.8 & 279.3 & 302.0 & $\>\>$58.6 \\
\hline
\end{tabular}
\caption{Values of the $\chi^2$ for the extended (before fitting) and refitted 
(after fitting) amplitudes assuming constraints on resonance parameters.}
\label{TableChi2-1}
\end{table}
Very big changes are seen in the $\chi^2_{Data}(S0)$ and $\chi^2_{DR}$ 
components. The big initial values are given mainly by the ill behavior 
of the extended amplitudes below $\approx$800~MeV which reflects their deficiency 
due to the crossing symmetry condition. 
This effect is also seen very well in Fig. \ref{FigReOld} below about 
600~MeV. The $\chi^2$ per number of degrees of freedom in this fit is 
$\chi^2/n.d.f.$ = 639.8/(494+78-31)=1.18.

Positions of the refitted poles changed notably for the $f_0(500)$, 
in particular those on sheets II and VII. The shift of the former was from 
$617 - i\,554$ to $463 - i\,296$~MeV which makes the new value well compatible 
with results based on ChPT and Roy-like equations~\cite{Caprini:2005zr}. 
Positions of poles of other resonances changed moderately by only few tens of per 
cent or less. The matching energy in the $S0$ wave became significantly smaller, 
$\sqrt{s_{00}} = 394$~MeV, which points to an improvement of the low-energy behavior 
of the multichannel amplitude above $\sqrt{s_{00}}$. Small values of the fitted 
background parameters indicate a small importance of the background part of the 
$S$-matrix, which is consistent with the spirit of our approach to the 
multichannel analysis of data presented in Ref. \cite{Yura2014}. The only small 
disruption of the method stems from a negative value of the elastic background 
parameter, $a_{11}= -0.0644$, suggesting that some part of description is still 
missing. The absolute value of $a_{11}$ is, however, small and the result is 
therefore acceptable. 

To improve the result we refitted the extended amplitudes again, removing now
the constraints imposed in the clusters on resonance poles for the $f_0(500)$, 
$f_0(980)$, $f_0(1500)$ and $\rho(770)$. In this case the number of free 
parameters was larger (43) but improvement of the $\chi^2$ makes this fit a bit 
better: $\chi^2/n.d.f.$ = 605.5/(494+78-43)=1.14.
Results of fitting with unconstrained resonance parameters are shown in 
Tables~\ref{TableChi2-2}--\ref{par_bgr_P1}.
%
%
\begin{table}[h!]
\hspace{.0cm}
\begin{tabular}{lcccc}  
\hline
& \multicolumn{1}{c}{$\chi^2$} & \multicolumn{1}{c}{$\chi^2_{Data}(S0)$} & 
\multicolumn{1}{c}{$\chi^2_{Data}(P1)$} & \multicolumn{1}{c}{$\chi^2_{DR}$} \\
\hline
extended  & ~1122.5  & 339.4 & 305.1 & 478.0 \\
refitted & ~$\>\>$605.5 & 269.0 & 300.9 & $\>\>$35.6 \\
\hline
\end{tabular}
\caption{Values of the $\chi^2$ for the extended (before fitting) and refitted 
(after fitting) amplitudes without the constraints on resonance parameters.}
\label{TableChi2-2}
\end{table}
Table \ref{TableChi2-2} shows the components of the total $\chi^2$ 
before and after fitting with unconstrained parameters of the resonances. 
An appreciable improvement with respect to the previous result 
is observed in the $\chi^2_{DR}$: $59\rightarrow 36$. This suggests that 
disabling the compactness of the resonance clusters allows the amplitudes 
to better conform the crossing symmetry condition driven by the GKPY equations.

%
%
\begin{table}[ht]
\begin{tabular}{crrr}
\hline
Sheet  &           & ~~extended~~      & ~~refitted~~\\
\hline
 & \multicolumn{3}{c}{$f_0(500)$}\\
II  & $E_r$        &  $616.5 $    & $455.9 $ \\ 
    & $\Gamma_r/2$ &  $554.0 $     & $295.4  $ \\ 
III & $E_r$        &  $621.8 $   & $826.4 $ \\ 
    & $\Gamma_r/2$ &  $554.0 $     & $168.4  $ \\ 
VI  & $E_r$        &  $598.3 $    & $709.4 $ \\ 
    & $\Gamma_r/2$ &  $554.0 $     & $185.3  $ \\ 
VII & $E_r$        &  $593.0 $     & $243.2 $ \\ 
    & $\Gamma_r/2$ &  $554.0 $     & $2093.9  $ \\ 
\hline
 & \multicolumn{3}{c}{$f_0(980)$}\\
II  & $E_r$        &  $1009.2 $  & $998.8 $ \\ 
    & $\Gamma_r/2$ &  $31.3 $    & $\,\,\,23.4  $ \\ 
III & $E_r$        &  $985.8 $   & $965.9 $ \\ 
    & $\Gamma_r/2$ &  $58.0 $    & $\,\,\,21.9  $ \\ 
\hline
 & \multicolumn{3}{c}{$f_0(1500)$}\\
II  & $E_r$        &  $1498.3 $   & $1441.7 $ \\ 
    & $\Gamma_r/2$ &  $198.8 $    & $164.1   $ \\ 
III & $E_r$        &  $1502.4 $   & $1514.4 $ \\ 
    & $\Gamma_r/2$ &  $236.8 $    & $97.12   $ \\ 
IV  & $E_r$        &  $1498.3 $   & $1468.5 $ \\ 
    & $\Gamma_r/2$ &  $193.0 $      & $145.5   $ \\ 
V   & $E_r$        &  $1498.3 $   & $1435.3 $ \\ 
    & $\Gamma_r/2$ &  $198.8 $    & $144.2   $ \\     
VI  & $E_r$        &  $1494.6 $   & $1507.0 $ \\ 
    & $\Gamma_r/2$ &  $194.0 $       & $162.8   $ \\ 
VII & $E_r$        &  $1498.3 $   & $1493.3  $ \\ 
    & $\Gamma_r/2$ &  $193.0 $      & $172.7   $ \\ 
\hline
\end{tabular}
\caption{Positions of the poles in the $S0$ amplitude before (extended) 
and after (refitted) fitting with unconstrained resonance parameters. 
Real and imaginary parts of the poles are given by 
$\sqrt{s_r}=E_r -i\Gamma_r/2$ in MeV.}
\label{par_poles_S0}
\end{table}
%
%
\begin{table}[ht]
\begin{tabular}{ccrr}
\hline
Sheet  &  & ~~extended~~  & ~~refitted~~\\
\hline
 & \multicolumn{3}{c}{$\rho(770)$}\\
II  & $E_r$        &  $766.4 $   & $765.1 $ \\ 
    & $\Gamma_r/2$ &  $72.4  $   & $73.2  $ \\ 
III & $E_r$        &  $766.4 $   & $915.0 $ \\ 
    & $\Gamma_r/2$ &  $72.4  $   & $23.3  $ \\ 
VI  & $E_r$        &  $766.4 $   & $0.4$ \\ 
    & $\Gamma_r/2$ &  $72.4  $   & $2.1  $ \\ 
VII & $E_r$        &  $766.4 $   & $1059.2 $ \\ 
    & $\Gamma_r/2$ &  $72.4  8$   & $0.01  $ \\
\hline
\end{tabular}
\caption{Positions of the poles in the $P1$ amplitude before and after fitting 
with unconstrained resonance parameters. Real and imaginary parts of the 
poles are given by $\sqrt{s_r}=E_r -i\Gamma_r/2$ in MeV.}
\label{par_poles_P1}
\end{table}

\begin{figure}[h!]
\centerline{\includegraphics[angle=0,scale=0.45]{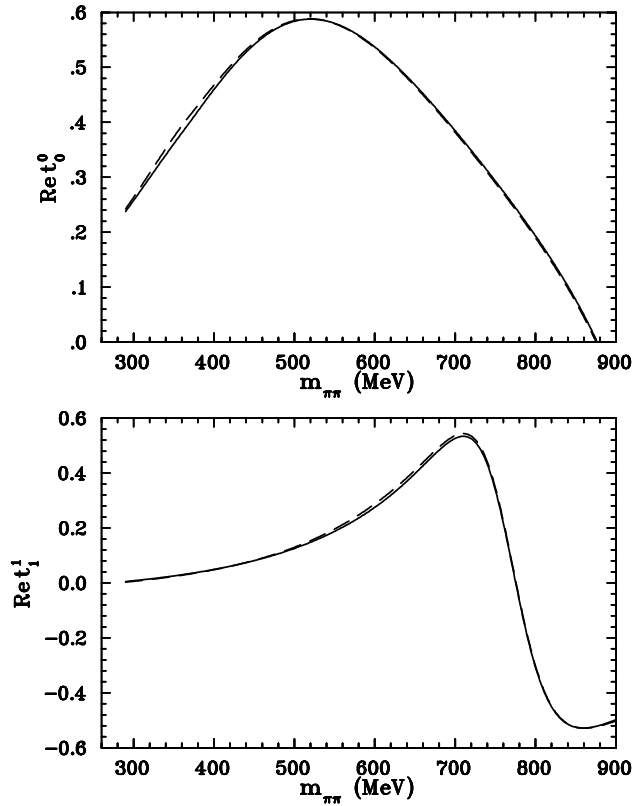}}
\caption{Input (dashed line) and output (solid line) real parts of the 
$S0$ (upper figure) and the $P1$ (lower figure) amplitudes after fitting.}
\label{FigReNew}
\end{figure}

Tables \ref{par_poles_S0} and \ref{par_poles_P1} show positions of the poles 
in the $S0$ and $P1$ amplitudes before and after fitting with unconstrained 
resonance parameters.
Significant change of the pole positions is apparent for the \sig meson.  
The dominant pole on the sheet II, which produces the biggest part of the phase 
shifts below 1~GeV, was now shifted by $161 + i\,259$~MeV towards the value 
recommended by PDG Tables 2012~\cite{PDGTables2012}. 
It is shown in Fig.~\ref{fig:InOut.D0} where the original pole of the 
multichannel amplitude \cite{Yura2010} is denoted by ``Old". 
Before fitting to the DR that pole was located even behind the range proposed 
by Particle Data Group before 2012 (denoted in the figure by 
``PDG2010") \cite{PDGTables2010}. After fitting, the new value 
$455.9\pm 8 -i\,295.4\pm 5$~MeV for the input amplitude and 
$449.3\pm 14 -i\,288.7\pm 14$~MeV for the output one locate very 
close to the center of the new - much smaller range given by the 
PDG2012 \cite{PDGTables2012} and is in a good agreement, e.g. with the 
value $441^{+16}_{-8}-i\,272^{+9}_{-13}$~MeV from~\cite{Caprini:2005zr}.
Large changes are observed also for the other pole positions, especially for 
the $\sigma$ meson. The quite small difference between $\sigma$ pole positions 
in the input and the output amplitudes is, as one can see in Fig. \ref{FigReNew}, 
due to a good agreement between them after fitting to the data and the DR.
In the whole text and in the tables we present position of the poles only for 
the input amplitudes.

The parameters of background changed moderately as it is seen in Tables \ref{par_bgr_S0} 
and \ref{par_bgr_P1}. Similarly as in the previous fit some of the parameters acquired 
negative values, e.g. $a_{11}$, $a_{1\sigma}$, $b_{1\eta}$ and $b$, disrupting a bit 
the philosophy of the multichannel approach in Ref. \cite{Yura2010}. 
However, as the absolute values are small this result is still acceptable. 
Value of the matching energy in the $S0$ wave is smaller than in the previous fit 
suggesting that in this case the multichannel amplitude above this energy is more 
flexible and can better accommodate requirements of the GKPY equations. 
%
%
\begin{table}[h!]
\hspace{.0cm}
\begin{tabular}{ccc}  
\hline
Parameter~ & \multicolumn{2}{c}{~value in the amplitude} \\
& extended & refitted \\
\hline
$a_{11}$      &  0.0124 & -0.0596 \\
$a_{1\sigma}$ &  0.0    & -0.1299 \\
$a_{1v}$      &  0.1004 &  0.1965 \\
$a_{1\eta}$   & -0.0606 &  0.0099 \\
$b_{1\sigma}$ &  0.0    & 0.0069 \\
$b_{1v}$      &  0.0469 &  0.1011 \\
$b_{1\eta}$   &  0.0    & -0.0052 \\
$\sqrt{s_{00}}$      &  525.7  &  382.2 \\
\hline
 \end{tabular}
\caption{Values of the background parameters and the matching energy $\sqrt{s_{00}}$ 
(in MeV) for the $S0$ wave before and after fitting with unconstrained resonance 
parameters.}
\label{par_bgr_S0}
\end{table}
%
%
\begin{table}[h!]
\hspace{.0cm}
\begin{tabular}{ccc}  
\hline
Parameter~ & \multicolumn{2}{c}{~value in the amplitude} \\
& extended & refitted \\
\hline
$a$      & -0.2851 & -0.34459 \\
$b$      & 0.00011 & -0.00020 \\
$\sqrt{s_{01}}$ &  643.6  &  635.4  \\
\hline
 \end{tabular}
\caption{Values of the background parameters and the matching energy 
$\sqrt{s_{01}}$ (in MeV) for the $P1$ wave before and after fitting 
with unconstrained resonance parameters.}
\label{par_bgr_P1}
\end{table}

Figure \ref{FigReNew}  presents a comparison of the input and output 
refitted $S0$ and $P1$ amplitudes. Comparison with Fig. \ref{FigReOld} 
shows how much the input and output amplitudes have changed.
The difference between them has significantly diminished what allows us 
to conclude that now the \pipi amplitudes fulfill the crossing symmetry 
really much better than before the fitting.
%
%

%
%
 It is desirable and instructive to check a stability of the results
 of analysis for the position of the $\sigma$ pole with respect to
 the definition of $\chi^2_{DR}$ in Eq. (\ref{EqChi2DR}).
 To this end several additional fits with various errors and numbers
 of chosen points for the $\chi^2_{DR}$ have been performed.
 The errors were chosen 2, 3 and 4 times smaller or larger than 0.01
 corresponding to presented in Table \ref{TableChi2-1} results, and
 the number of points, $N_{DR}$, was the same (26) or two times 
 bigger. The uncertainties of the real and imaginary parts of the 
 $\sigma$ pole positions resulting from these changes proved to be 
 much smaller than the error 14 MeV found in our best fit. They were: 
 3.4 MeV for the real part and 3.3 MeV for the imaginary one. Values 
 of the $\chi^2$ for the experimental data used in the fits varied 
 between 1.08 and 1.27 per degree of freedom. Clearly, variations of 
 the $\chi^2_{DR}$ were larger but they followed changes in the error 
 and the number of points.
 
 Summarizing this check, one has to emphasize the clear stability of
 the results presented in the analysis and their significant independence
 from the chosen definition of $\chi^2$.
 Nevertheless in other similar analyses, particularly in those made with
 other experimental data leading to amplitudes deviating significantly
 from those allowed by the crossing symmetry condition (see for 
 example Section VII in \cite{GKPY}), one has to always choose a definition 
 of the $\chi^2_{DR}$
 giving magnitudes of the same order as the $\chi^2_{Data}$.
%
%

As the most significant effect of refining the amplitudes was found to be 
the shift of the $\sigma$ pole, it is also desirable to show whether this effect 
depends on specific properties of the $S0$ amplitude or on the imposed crossing 
symmetry condition. For this purpose we have constructed a new version 
of the $S0$ input amplitude removing also imperfection of the old one 
which we found during our calculations, particularly a small violation 
of two-body unitarity around 1.2~GeV. Parameters of the new amplitude 
were not fitted to the DR but only to the same data and following the 
same method as in Ref.~\cite{Yura2010}. 
The result of fitting is a bit worse than that in \cite{Yura2010}, 
$\chi^2$ = 302.3/(294-38)= 1.18 (1.11 in Ref.~\cite{Yura2010}), 
but the new amplitude is without the shortcoming due to unitarity. 
Moreover, in this new amplitude the $\sigma$ pole on sheet II at 
$562.9 -i\,417.1$~MeV (denoted in Fig. \ref{fig:InOut.D0} by ``New") 
is nearer to the PDG 2012 region than before.
Especially the magnitude of imaginary part is smaller by 137~MeV. 
In Table~\ref{newS0} we show positions of all poles of the new $S0$ 
amplitude on the Riemann surface. 
%
%
\begin{table*}[thb] 
\caption{The pole clusters for the new refitted $S0$ wave amplitude. 
$\!\sqrt{s_r}={\rm E}_r-i\Gamma_r/2\!$~ in MeV is given.} 
\label{newS0} 
\begin{center} 
\begin{ruledtabular} 
{\small 
\begin{tabular}{ccccccccc} 
Sheet   &    &  II  &  III  &  IV  &  V  &  VI  &  VII  &  VIII \\ 
\hline 
{$f_0(500)$}~~  &  {$\!{\rm E}_r\!$}~~ & $562.9 \pm 4.9$ & $594.7 \pm 7.8$ & 
   &     & $615.1 \pm 39.6$ & $583.3 \pm 17.4$ & \\ 
                & {$\!\Gamma_r/2\!$}~~ & $417.1 \pm 14.9$ & $417.1 \pm 14.9$ &  
   &     & $417.1 \pm 14.9$ & $417.1 \pm 14.9$ & \\ 
\hline 
{$f_0(980)$}~~  & {$\!{\rm E}_r\!$}~~ & $1007.6 \pm 3.0$ & $984.5 \pm 5.4$ & 
  &  &  &  & \\ 
                & {$\!\Gamma_r/2\!$}~~ & $29.4 \pm 1.5$ & $55.1 \pm 3.4$ & 
&   &    &   &  \\ 
\hline 
{$f_0 (1370)$}~~& {$\!{\rm E}_r\!$}~~ &   & $1397.0 \pm 6.3$ & $1397.0 \pm 6.3$ & 
$1277.0 \pm 50.2$ & $1277.0 \pm 50.2$ &  &  \\ 
                & {$\!\Gamma_r/2\!$}~~ &  & $216.5 \pm 13.2$ & $180.1 \pm 11.1$ & 
$180.1 \pm 11.1$ & $216.5 \pm 13.2$ &   &   \\ 
\hline 
{$f_0(1500)$}~~ & {$\!{\rm E}_r\!$}~~ & $1496.6 \pm 4.7$  & $1502.7 \pm 3.8$  &
$1496.6 \pm 4.7$ & $1496.6 \pm 4.7$ & $1502.1 \pm 2.9$ & $1496.6 \pm 4.7$ &  \\ 
                & {$\!\Gamma_r/2\!$}~~ & $203.5 \pm 4.5$ & $223.9 \pm 5.2$  & 
$198.7 \pm 7.4$  & $203.5 \pm 4.5$ & $187.9 \pm 4.0$ & $198.7 \pm 7.4$  &  \\ 
\hline 
{$f_0 (1710)$}~~& {$\!{\rm E}_r\!$}~~ &   &   &   & 
$1736.0 \pm 26.6$  & $1736.0 \pm 26.6$ & $1736.0 \pm 26.6$ & $1736.0 \pm 26.6$ \\ 
                & {$\!\Gamma_r/2\!$}~~ &  &  &  & $108.3 \pm 16.6$ & $85.5 \pm 11.8$ & 
$ 43.3 \pm 14.1$ & $66.1 \pm 13.8$ 
\end{tabular} 
} 
\end{ruledtabular} 
\end{center} 
\end{table*} 

The new background parameters in Eq.(\ref{bgr_s-wave}) 
are: $a_{11} =-0.0131\pm0.0019$, 
$a_{1\sigma} =0$, $a_{1v} =0.046\pm0.011$, $a_{1\eta} =-0.0302\pm0.0025$, 
$b_{11} =b_{1\sigma} =b_{1\eta} =0$, $b_{1v} =0.0573\pm0.0071$, 
$a_{21} =-3.5456\pm0.0188$, $a_{2\sigma} =0.6778\pm0.0850$, $a_{2v} =-4.28\pm0.44$, 
$a_{2\eta} =-0.1857\pm0.0304$, $b_{21} =0$, $b_{2\sigma} =0.6499\pm0.0886$, 
$b_{2v} =1.49\pm1.74$, $b_{2\eta} = 0.1011\pm0.0265$, $b_{31} =0.3825\pm0.0412$, 
$b_{3\sigma} =b_{3v} =b_{3\eta} =0$,
$s_\sigma=1.638~{\rm GeV}^2$, $s_v=2.126~{\rm GeV}^2$. 
In the new extended $S0$ amplitude the matching energy is $\sqrt{s_{00}}=406.5$~MeV.
Note that, similarly as in our previous fits starting with the ``old" amplitude the 
parameter $a_{11}$ is negative.

We used this new extended $S0$ amplitude in the DR analysis keeping all other 
ingredients unchanged as in our previous two fits. During the fitting we let 
all poles of resonances to vary independently as we did it in our second fit 
with the old extended $S0$ amplitude. 
Values of $\chi^2$ are shown in Table~\ref{TableChi2-3}. 
%
%
\begin{table}[h!]
\hspace{.0cm}
\begin{tabular}{lcccc}  
\hline
& \multicolumn{1}{c}{$\chi^2$} & \multicolumn{1}{c}{$\chi^2_{Data}(S0)$} & 
\multicolumn{1}{c}{$\chi^2_{Data}(P1)$} & \multicolumn{1}{c}{$\chi^2_{DR}$} \\
\hline
extended  & ~1085.4  & 302.3 & 305.1 & 478.0 \\
refitted & ~$\>\>$607.8 & 274.2 & 293.9 & $\>\>$39.7 \\
\hline
\end{tabular}
\caption{Values of the $\chi^2$ for the new extended (before fitting) 
and refitted (after fitting) amplitudes.}
\label{TableChi2-3}
\end{table}
The fit is equally good as the previous one: 
$\chi^2/n.d.f.$ = 607.8/(494+78-43)=1.15. The $\sigma$ pole on sheet II 
is now at $459.0\pm 8 -i\,292.3\pm 5 $~MeV for the input amplitude and at 
$445.2 \pm 14 - i\,296.4 \pm 14$~MeV for the output one which is very well 
consistent with the previous results (see Fig.~\ref{fig:InOut.D0}). 
This suggests that the $\sigma$ pole position is prescribed rather by the 
GKPY equations than by the data or a structure of the $S0$ amplitude. 

In order to know how released scattering length parameters, i.e., $a^0_0$ and  
$a^1_1$ affect the value of total $\chi^2$ for the $S0$ and $P1$ wave, 
we made them free and performed fit again. The total $\chi^2/n.d.f.$ slightly 
changed from 1.145 to 1.137 which is neglectable. 
Values of the fitted scattering lengths were:
$a^0_0=0.224$ $m_\pi^{-1}$, $a^1_1=0.0340$~$m_\pi^{-3}$.
%
%
\section{Uniqueness of the results}
\label{proof}

Uniqueness of the results on the \sig pole position presented in the previous 
section can be proved using purely mathematical arguments.
One should start, however, with the arguments for uniqueness of results obtained 
in the dispersive data analysis presented in Ref. \cite{GKPY,PreciseDet}.
These have been received without any model assumptions about specific energy 
dependence of the $\pi\pi$ amplitudes. 
Another although similar analysis performed for the Roy equations has used 
two assumptions for values of the $S0$ wave amplitude at 800~MeV and at the  
$\pi\pi$ threshold \cite{A4}.
In accordance with the method described in Ref. \cite{Wanders2000} due to these 
two boundary conditions the authors found the unique analytical solution of the 
Roy equations below 800~MeV. 
The position of the $\sigma$ pole obtained in \cite{GKPY,PreciseDet} differs 
by less then one standard deviation from that received in \cite{A4} what 
ensures in correctness and uniqueness of the results found in analyses \cite{GKPY} 
and \cite{PreciseDet} using the GKPY equations.

\begin{figure}[h!]
\centerline{\includegraphics[angle=0,scale=0.45]{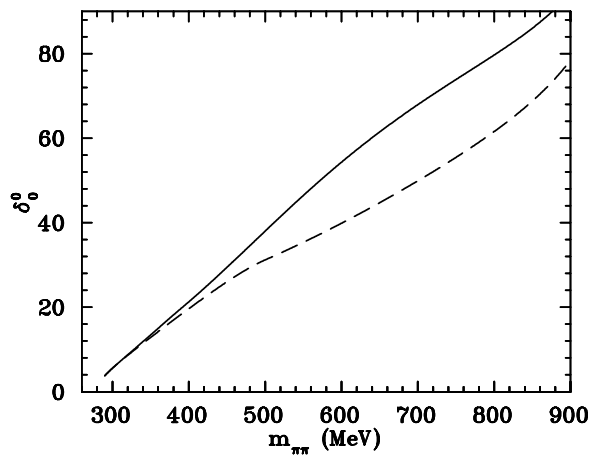}\vspace{-50mm}}
\caption{Effective two-pion mass dependence of the phase shifts corresponding 
to the extended (dashed line) and refitted (solid line) ``Old" amplitudes.}
\label{FigPhaseOldNew}
\end{figure}

Uniqueness of the new position of the \sig pole and of its movement found 
in our analysis after fit to the DR can be easily proved using only two simple arguments:
trigonometric relations satisfied by the \pipi amplitudes and constraints given 
by the crossing symmetry condition. Looking at Fig. \ref{FigPhaseOldNew} 
with the $S0$ phase shifts corresponding to the extended and the refitted 
``Old" amplitudes one can observe significant differences between them 
especially from 500 to 800~MeV. One should also notice their completely 
different curvatures in this region caused by very different positions of 
the $\sigma$ pole presented in the previous Section.
%
%

\begin{figure}[h!]
\centerline{\includegraphics[angle=90,scale=0.35]{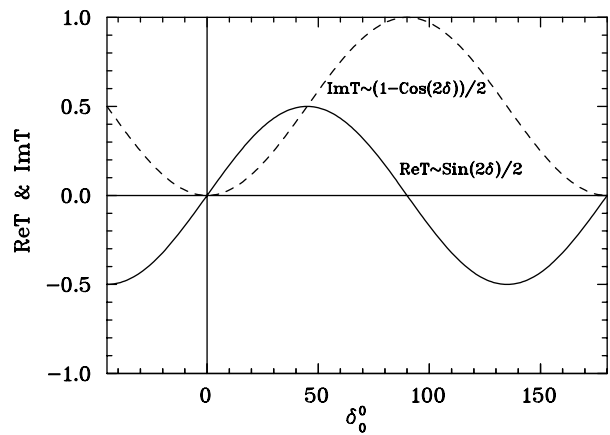}\vspace{-10mm}}
\caption{Dependence of the real and imaginary part of the amplitude on the phase shifts.}
\label{FigReImTrygo}
\end{figure}

\begin{figure}[h!]
\centerline{\includegraphics[angle=90,scale=0.35]{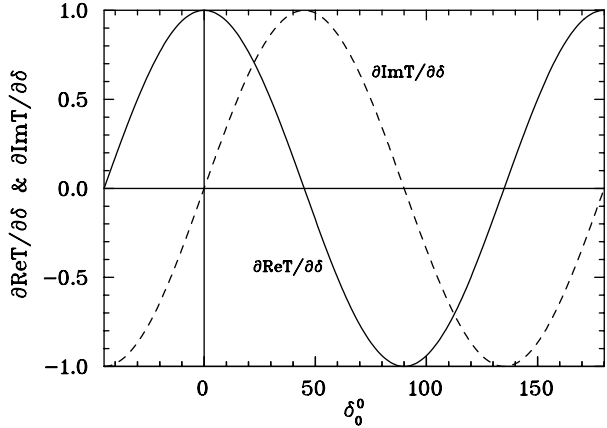}\vspace{-10mm}}
\caption{Gradient of the real and imaginary part of the amplitude as a 
function of the phase shifts.}
\label{FigDtReDtImTrygo}
\end{figure}

\begin{figure}[h!]
\centerline{\includegraphics[angle=90,scale=0.37]{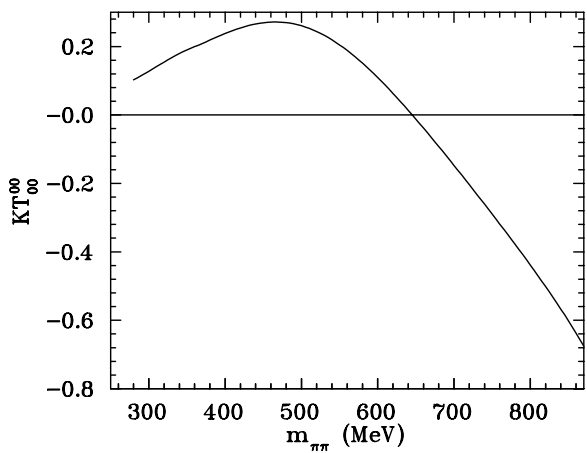}\vspace{-10mm}}
\caption{Effective two pion mass dependence of the $KT^{00}_{00}$ term.}
\label{FigK0000}
\end{figure}

Real parts of the $S0$ input and output amplitudes presented in  
Fig.~\ref{FigReOld} correspond to the extended ``Old" amplitude.
In order to diminish difference between them one could intuitively think 
on a shifting down and up the real parts of the input amplitude below and 
above about 650~MeV respectively.
The dependence of the real part of the amplitude on the phase shifts is presented 
in Fig.~\ref{FigReImTrygo} and indicates that it would lead to decrease
of the phase shifts in whole $m_{\pi\pi}$ region below about 900~MeV.
%
%
As is, however, seen on this figure it would reduce also the imaginary part of 
the amplitude. Comparison of the gradients of the imaginary and real parts 
of the amplitude in Fig.~\ref{FigDtReDtImTrygo} (i.e., of the input and the output 
functions in Eq. (\ref{EqGKPY}), respectively) shows that for the phase shifts 
between around 22 and 112 degrees, i.e., within the range that we are interested 
in (see for example Fig.~\ref{FigBydzNoDR}), just the imaginary part changes faster 
than the real one.
%
%
Figure \ref{FigK0000} presents energy dependence of the kernel part $KT^{00}_{00}$ being 
dominant in the full output amplitude ${\mbox{Re } \bar t_{0}^{0}(s)}$ in Eq. (\ref{EqGKPY})
\begin{equation}
KT^{00}_{00}(s) = \hspace{0.05cm}-\hspace{-0.49cm}\displaystyle \int \limits_{4m_{\pi}^2}^{\Lambda}\hspace{-0.05cm} ds'
     K_{00}^{00}(s,s')\, { \mbox{Im }t_{0}^{0(IN)}}(s')
\label{EqKT0000}
\end{equation}
where the parameter $\Lambda = (1.42\,\, \mbox{GeV})^2$ is upper limit of integration 
for the phenomenologically parametrized amplitudes (for details see Section II C of \cite{GKPY}).
%
%
Characteristic and important feature is the positive value of the $KT^{00}_{00}$ 
below about 650~MeV and its negative value above this energy.
Smooth and monotonic energy dependence of the input amplitude 
${\mbox{Im }t_{0}^{0}}(s')$ below about 800~MeV, caused by smooth behavior 
of the phase shifts, guarantees that such shape is produced by the kernel 
$K^{00}_{00}(s)$ in the Eqs. (\ref{EqGKPY}) and (\ref{EqKT0000}).
This has been checked for different parametrizations of the phase shifts 
below 1~GeV.

Taking into account the bigger gradient of the ${\mbox{Im }t_{0}^{0}}(s')$
than that of the ${\mbox{Re }t_{0}^{0}}(s')$
and the shape of the $KT^{00}_{00}$, one can conclude that the intuitively 
expected and mentioned above decrease of the phase shifts would cause faster 
decrease of the output amplitude than decrease of the input one below around 
650~MeV and faster increase above this energy. It means that the input 
amplitude could not catch up the escaping output one. Therefore in order to 
diminish distance between the real parts of the input and output amplitudes
after fitting, the phase shifts must not decrease but increase below about 800~MeV.
Then, finally, for some values of the increasing phase shifts, the input and 
output real parts can almost overlap what one can observe in Fig.~\ref{FigReNew} 
for the refitted amplitudes.
Comparing the input and output real parts for the extended and refitted amplitudes 
in Figs.~\ref{FigReOld} and \ref{FigReNew} one can easily check that below 650~MeV 
the input amplitude really increased and above  this energy really decreased after 
fitting to the DR.

This pure mathematical analysis proofs that the increase of the phase shifts 
followed by change of the curvature after fitting, seen in Fig.~\ref{FigPhaseOldNew}, 
is natural and unique consequence of the trigonometry 
relations satisfied by the input and output amplitudes and of the energy dependence 
of the kernels given by the crossing symmetry condition. Analyzing the analytical 
structure of the amplitude in the complex energy plane one can conclude that such 
modification of the phase shifts can be produced only by the \sig pole on the 
IInd Riemann sheet moving toward the physical and the imaginary axis, i.e. 
by a narrower and lighter \sig meson. 
This we have illustrated in Fig.~\ref{fig:InOut.D0}.

%
%
\section{Conclusions}

Very effective and simple way of modifying the amplitudes of the \pipi interactions 
in the $S0$ and $P1$ waves has been presented.
These modifications do not change mathematical structure 
of the original multichannel amplitudes (analytical properties on the Riemann surface)  
but only fit some of their parameters to the dispersion relations with 
the imposed crossing symmetry condition.
After these modifications the amplitudes fulfill this symmetry from the \pipi 
threshold to around 1.1~GeV and describe very well experimental data below about 
1.8~GeV in all considered channels, e.g., $\pi\pi$, $K \bar K$, and $\eta\eta'$.
In the case of the amplitudes considered here, apart of finding new values of 
some of their parameters, the only important modification was introduction 
of the new part describing the near threshold region. The most important 
consequence of the fit to the dispersion relations was a significant 
movement of the \sig pole by several hundred MeV, i.e. by values comparable 
with its final mass and width.
The new $\sigma$-pole position agrees very well with the value accepted
in the Particle Data Tables in 2012 \cite{PDGTables2012}. 
The amplitudes, refitted using the presented method, can now be used as 
representative for the modeling of the \pipi interactions what was impossible 
for the original ones with significantly too massive and too wide \sig meson.
Another interesting result of refitting the multichannel amplitudes 
is a marked change of positions of the $f_0(1500)$ poles which are located 
well above the inelastic threshold. On the contrary imposing the crossing 
symmetry constraint practically did not affect the parameters of $\rho (770)$.

The method described here and presented example can be very helpful in refitting 
of other existing amplitudes which have similar problems as the original 
amplitudes analyzed here.

\vspace*{0.3cm}

{\bf ACKNOWLEDGMENT} The authors are grateful to Yu.S. Surovtsev for useful 
discussions. This work has been funded by the Polish National Science 
Center (NCN) Grant No. DEC-2013/09/B/ST2/04382 and was partially supported by 
the Grant Agency of the Czech Republic under Grant No. P203/12/2126.
%

\end{document}